\documentclass[fleqn,twoside]{article}
\usepackage{espcrc2}

\usepackage{graphicx}
\usepackage[figuresright]{rotating}

\newcommand{\AmS}{{\protect\the\textfont2
  A\kern-.1667em\lower.5ex\hbox{M}\kern-.125emS}}

\title{Towards measuring the $\eta'$ mass with $N_f=2+1$ staggered fermions}

\author{Eric B. Gregory\address[LIV]{Theoretical Physics Division,
Department of Mathematical Sciences,\\University of Liverpool,
Liverpool, L69-7ZL, United Kingdom}, Alan Irving\addressmark[LIV], Craig McNeile\addressmark[LIV], Steven Miller\addressmark[LIV], Zbyszek Sroczynski\addressmark[LIV]
}
\newcommand{\me}{\mathrm{e}}
\def\Dslash{\mathop{\not\!\! D}}
       
\begin{document}

\begin{abstract}
Because the propagators of flavor-singlet states incorporate disconnected
diagrams, they are uniquely sensitive to any differences in the actions
governing sea and valence fermions on the lattice. As such, they present an
important test of the validity of  the ``fourth-root trick'' in the staggered
fermion formulation. The pseudoscalar's relationship to topological charge
also makes it of theoretical interest. We present preliminary results from
our measurements of flavor-singlet pseudoscalar mesons on 2+1 flavor Asqtad
lattices,  and discuss some strategies for improving the signal-to-error ratio
of disconnected singlet correlators.
\vspace{1pc}
\end{abstract}

\maketitle

\section{INTRODUCTION}
The mass of the pseudoscalar singlet meson, the $\eta'$, has long been the 
focus of theoretical study. It is more than 800MeV heavier than the pion, 
a difference that has been attributed to disconnected quark 
loops\cite{Witten:1979vv,Veneziano:1979ec}, which are not
present in the propagators of non-singlet states. Lattice 
calculations using quenched Wilson fermions \cite{Itoh:1987iy}, 
$N_f=2$ Wilson \cite{McNeile:2000hf,Struckmann:2000bt,Lesk:2002gd,Schilling:2004kg,Allton:2004qq},
and $N_f=2$ staggered fermions 
\cite{Venkataraman:1997xi,Kogut:1998rh}, lend credence to this 
idea.
  
In the real world, 
there appears to be significant mixing between the singlet 
and the non-singlet mesons to form the $\eta$ and $\eta'$ mass eigenstates. 
For lattice calculations to explain
the observed pseudoscalar meson spectrum, it is therefore important to
consider the effect of strange quarks as well as the two light flavors. In
this paper we describe the very preliminary stages of the our efforts to study
the $\eta$ and $\eta'$ mesons using $N_f=2+1$ flavors (two light degenerate flavors
and one heavier ``strange'' flavor) of staggered quarks. We also note that,
as the disconnected diagrams inherent in the singlet propagator are sensitive
to the sea quarks, flavor singlet studies may highlight any problems which
the ``fourth-root trick'' introduces into the staggered sea.

\section{THEORETICAL BACKGROUND}

The difference between singlet and non-singlet pseudoscalar mesons is evident 
by looking at the expression for the singlet propagator. The general 
pseudoscalar $\overline{\psi}\gamma_5\psi$ propagator is:
\begin{equation}
\Bigg\langle\left[\sum_{i}^{N_f}\overline{q}_i(x')\gamma_5 q_i(x')\right]^\dagger\sum_{j}^{N_f}\overline{q}_j(x)\gamma_5q_j(x)\Bigg\rangle.
\end{equation}
This expression contains terms of two types:
\begin{itemize}
\item 
For $N_f$ flavors, there are $N_f$ terms with contractions 
of fields from $x$ to $x'$. :
\begin{equation}
\Bigg\langle\left[\sum_{i}\right.\overbrace{\overline{q}_i(x') \gamma_5 \overbrace{q_i(x')\left.\Bigg]\right.^\dagger\sum_{j}\overline{q}_j}(x)\gamma_5 q_j}(x)
\Bigg\rangle
\end{equation}
These terms appear in the propagators of both the singlet and non-singlet 
states.
\item $N_f^2$ terms from contractions of fields at the same space-time point\\
\begin{equation}
\Bigg\langle\left[\sum_{i}\overbrace{\overline{q}_i(x')\gamma_5q_i}(x')\right]^\dagger\sum_{j}\overbrace{\overline{q}_j(x)\gamma_5q_j}(x)\Bigg\rangle
\end{equation}
These terms are present {\em only} in the flavor-singlet and 
distinguish the $\eta'$ from the pion. They give 
rise to disconnected diagrams.
\end{itemize}

To understand how the disconnected terms can contribute to an increase in
the singlet mass relative to the non-singlet, it is instructive to 
expand the total singlet propagator in terms of pions with an effective
coupling of $-\mu^2$. The first term, the connected term, is the pion
propagator:
\begin{equation}
\label{G0}
G_0(p)=\frac{1}{p^2+m_\pi^2}.
\end{equation}
Following that is the disconnected piece, expanded as $-D(p)=\sum_{i=1}^\infty G_i(p)$,
where the $i$th term has $(i-1)$ {\em sea} quark loops (pion propagators):
\begin{eqnarray}
G_1(p)&=&-\frac{1}{p^2+m_\pi^2}\mu^2\frac{1}{p^2+m_\pi^2}\label{G1}\\
G_2(p)&=&\frac{1}{p^2+m_\pi^2}\mu^2\frac{1}{p^2+m_\pi^2}\mu^2\frac{1}{p^2+m_\pi^2}\\
&& ... \nonumber\\
G_i(p)&=&\frac{1}{p^2+m_\pi^2}\left[\frac{-\mu^2}{p^2+m_\pi^2}\right]^{i},
\end{eqnarray}
such that the entire geometric series sums to a propagator with the mass 
squared shifted by $\mu^2$:
\begin{eqnarray}
G(p)&=&\sum_{i=0}^\infty G_i(p)\nonumber \\
&=& \frac{1}{p^2+m_\pi^2}\sum_{i=0}^\infty
\left[\frac{-\mu^2}{p^2+m_\pi^2}\right]^{i}\nonumber\\
&=&\frac{1}{p^2+m_\pi^2+\mu^2}.
\end{eqnarray}
  
In addition to giving a perturbative explanation for the heavier mass of the 
$\eta'$, this expansion also illustrates the importance of sea quark loops
in the $\eta'$ propagator. 

\subsection{The $D/C$ ratio}

In Euclidean configuration space, the connected piece Eq. \ref{G0} should have 
exponential behavior:
\begin{equation}
C(t) \sim \me^{-m_\pi t},
\end{equation}
as should the total singlet propagator:
\begin{equation}
N_fC(t)-N_f^2D(t)\sim \me^{-m_\eta' t}.
\end{equation}
So we expect the ratio of the disconnected correlator to connected correlator
to behave as
\begin{equation}
\label{full_ratio}
\frac{N_f^2D(t)}{N_fC(t)}= 1-A\me^{-(m_{\eta'}-m_\pi) t}.
\end{equation}

In a world without sea quarks ---  described by quenched lattice simulations 
--- the momentum space expansion would terminate after $G_1(p)$ (Eq \ref{G1}).
In such a case the behavior of the $D/C$ ratio would instead 
be \cite{Bernard:1992mk,Bernard:1993sv}:
\begin{equation}
\label{quenched_ratio}
\frac{D(t)}{C(t)}= A-Bt.
\end{equation}
The $D/C$ ratio is a useful tool to highlight any misbehavior on the part
of the sea quarks in a lattice simulation. In simulations involving 
Kogut-Suskind staggered fermions the four native tastes of sea quark are 
reduced to a single flavor by means of the ``fourth-root trick''. It is
possible that replacing the determinant of the fermion matrix by the
fourth-root of the determinant may introduce non-local terms in the action 
governing the fermion sea. Measuring the $D/C$ ratio and looking for deviations
from the form of Eq. \ref{full_ratio} would be a useful way to spot any 
inconsistencies between the actions governing the sea and valence quarks.

One such inconsistency, albeit a mild one, is already known. The fourth-root
trick is applied only to the sea quarks, reducing the number of flavors 
orbiting a sea quark loop by a factor of $1/4$. Valence quark loops still
have the native four tastes of staggered fermions orbiting. As connected 
correlators have a single valence loop and disconnected correlators have
two valence quark loops, the ratio $D/C$ is naively too large by a factor of 
four for staggered fermions. In the numerical discussion following, we 
rescale implicitly all 
disconnected correlators by $1/4$ to correct for this.

\section{SIMULATION \& MEASUREMENT}
\subsection{Gauge configurations}

Table \ref{config_table} lists the gauge configuration ensembles on which
we have to date measured pseudoscalar singlet connected and disconnected
correlators. The $16^3\times32$ configs were used to test our algorithms
and were generated at the University of Liverpool. The $20^3\times64$
configurations are ``coarse'' MILC lattices with $a\approx0.12$fm \cite{Bernard:2001av,Aubin:2004wf}.
All are produced with the ``Asqtad'' improved 
action \cite{Orginos:1998ue,Orginos:1999cr,Lepage:1998vj}.

\begin{table*}[htb]
\label{config_table}
\newcommand{\m}{\hphantom{$-$}}
\newcommand{\cc}[1]{\multicolumn{1}{c}{#1}}
\renewcommand{\tabcolsep}{2pc} 
\renewcommand{\arraystretch}{1.2} 
\begin{tabular}{@{}llllll}
\hline
$N_f$           & $10/g^2$ & $L^3\times T$ & $am_{\rm sea}$ & $am_{\rm val}$ & $N_{\rm configs}$ \\
\hline
\hline
0         &   8.0  & $16^3\times 32$  & ---  & 0.020 & 76 \\
2         &   7.2  & $16^3\times 32$  & 0.020 & 0.020 & 268 \\

\hline
\hline
0         &   8.00 & $20^3\times 64$  & ---  & 0.020 & $408^*$ \\
\hline
2         &   7.20 & $20^3\times 64$  & 0.020 & 0.020 & $547^*$ \\
\hline
2+1       &   6.76 & $20^3\times 64$  & 0.007, 0.05 & 0.007, 0.05 & 422\\
2+1       &   6.76 & $20^3\times 64$  & 0.010, 0.05 & 0.010, 0.05 & 644\\

\hline
\end{tabular}\\[2pt]
\caption{Ensembles used in pseudoscalar singlet measurement to date. * denotes
analysis is still in progress.}
\end{table*}

\subsection{Singlet operators}
Two different staggered meson operators can be used as a base for 
coupling to pseudoscalar taste-singlet states: 
$\gamma_4\gamma_5\otimes{\bf 1}$ and $\gamma_5\otimes{\bf 1}$. The former is
a three-link operator, with the quark and anti-quark sources located on 
opposite corners of a spatial cube, separated by three gauge links. The 
$\gamma_5\otimes{\bf 1}$ is a four-link operator, with the quark and anti-quark
sources situated on opposite corners of a hypercube, separated by four
gauge links. We measure connected and disconnected correlators on the 
configurations using both, effecting the separation between quark and 
anti-quark sources with covariant symmetric shifts.  In practice, however,
we use only the four-link $\gamma_5\otimes{\bf 1}$ operator for the analysis
described below, for the reason that its correlators come with no 
oscillating parity partner state. The  $\gamma_4\gamma_5\otimes{\bf 1}$ 
has as its partner the scalar ${\bf 1}\otimes\gamma_4\gamma_5$,
but the partner
of the  $\gamma_5\otimes{\bf 1}$ is $\gamma_4\otimes\gamma_4\gamma_5$, which is 
exotic. Hence the correlators of the $\gamma_5\otimes{\bf 1}$ are much easier 
to analyze.

\subsection{Measuring correlators}
We use the Chroma lattice QCD software system \cite{Edwards:2004sx} 
to measure connected and disconnected correlators. 
Measuring connected correlators is straightforward, particularly for 
pseudoscalars. Point sources are sufficient to obtain connected correlators
which are discernible nearly across the entire time-span of the lattice, 
despite decaying through many orders of magnitude. See Fig \ref{b676ccorr}
as an example.

Disconnected correlators are considerably more difficult to measure and are 
inherently noisy, as the correlation is communicated through gluons and sea 
quarks only. To measure a signal we use the now-standard method of 
volume-filling stochastic noise 
sources \cite{Bitar:1988bb,Fiebig:1990uh,Dong:1993pk,Kuramashi:1994aj,Eicker:1996gk}.

\begin{figure}[htb]
\rotatebox{270}{\includegraphics[width=15pc]{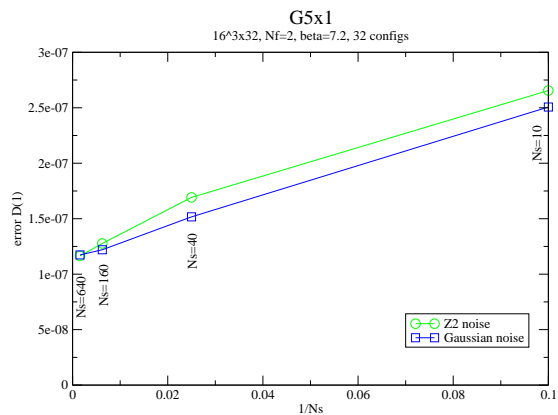}}
\caption{Noise dependence of the $t=1$ element of the $\gamma_5\otimes{\bf 1}$
disconnected correlator as a function of the number of noise sources for both 
$Z_2$ and Gaussian noise.}
\label{z2gauss}
\end{figure}

We tested both Gaussian and $Z_2$ noise sources on the two-flavor 
$16^3\times 32$ ensemble and found the statistical errors were consistently 
smaller for the Gaussian sources. Figure \ref{z2gauss} shows the dependence of
the error of the disconnected correlator on the number of sources for both
$Z_2$ and Gaussian noise.

In general one defines noise source vectors $\eta^i(x)$ on the entire 
lattice. The sources obey:
\begin{equation}
\lim_{N_S\rightarrow\infty}\frac{1}{N_S}\sum_{i=0}^{N_S} \eta^{i*}_x\eta^i_y = \delta_{xy},
\end{equation}
so that with a large number of sources $N_S$ we can solve $\phi_y=M^{-1}_{xy}\eta_y^i$ and then determine the loop operator
\begin{eqnarray}
\label{normal}
{\mathcal O}_{\gamma_5\otimes{\bf 1}}(t)=&\langle {\rm Tr}\Delta_{xy} M^{-1}_{yx}\rangle \nonumber\\
=&\frac{1}{N_S}\Bigg\langle\sum\limits_{i=0}^{N_S}\sum\limits_{{\bf x}}\eta_x^{i*}\Delta_{xy}\phi^i_y\Bigg\rangle_{x_0=t},
\end{eqnarray}
where the operator $\Delta_{xy}$ effects the four-link
shift and Kogut-Susskind phasing appropriate for the ${\gamma_5\otimes{\bf 1}}$
operator.

\subsection{Variance reduction schemes}
We have tested a variance reduction trick used by Venkataraman and 
Kilcup \cite{Venkataraman:1997xi}. Specifically, for the 
$\gamma_5\otimes{\bf 1}$ operator we can estimate 
$\langle {\rm Tr} \Delta_{xy}^{\gamma_5\otimes{\bf 1}} M^{-1}_{yx}\rangle$ as 
\begin{equation}
\label{kilcup}
\langle {\rm Tr} \Delta_{xy}^{\gamma_5\otimes{\bf 1}} M^{-1}_{yx}\rangle
=m\Big\langle\frac{1}{N_s}\sum_i\phi_x^{i*}\Delta_{xy}^{\gamma_5\otimes{\bf 1}}\phi^i_y\Big\rangle.
\end{equation}
We note that this expression differs from Eq. \ref{normal} by
$\langle\eta_x \Dslash (D^2+m^2)^{-1} \eta_y\rangle$. The staggered $\Dslash$ 
connects even and odd sites only, and $(M^\dagger M)^{-1}=(D^2+m^2)^{-1}$ 
connects only sites separated by an even number of links. So when $x$ and $y$ 
are 
separated by an even number of links this term is zero. This is the case for 
the four-link $\gamma_5\otimes{\bf 1}$, but of course is not for 
$\gamma_4\gamma_5\otimes{\bf 1}$. The advantage of this substitution is that
while both forms have the same expectation value, the form in 
Eq. \ref{kilcup} has a smaller variance.

We have also experimented with dilute noise sources.
A recent work by the TrinLat Collaboration has suggested that using dilute
stochastic sources can reduce the error of quantities measured with 
stochastic sources \cite{Foley:2005ac}. In this scheme
the stochastic source vectors are non-zero only on some subset of the
lattice, with the set of noise vectors designed such that the entire lattice 
is covered in the sum over sources. With staggered fermions several different 
dilution schemes are possible, e.g.: dilution by time-slice, color, 
site-parity, hypercube component, as well as the intersections of these 
dilution schemes. We have tested several of dilution schemes and mention
the results briefly in subsection  \ref{var_red}

\section{ANALYSIS \& RESULTS}
\subsection{$N_f=2+1$ flavor results}
\begin{figure}[tb]
\rotatebox{270}{\includegraphics[width=15pc]{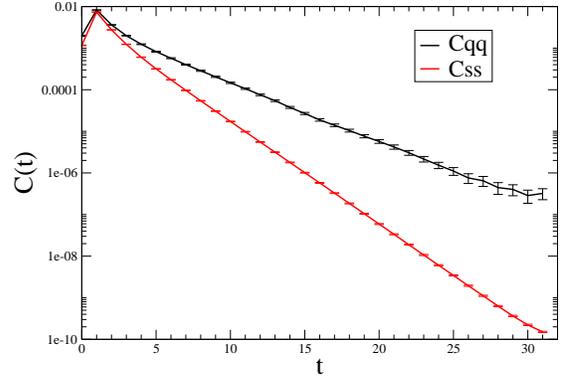}}
\caption{Connected correlators of the $\gamma_5\otimes{\bf 1}$ operator
for $N_f=2+1$ $\beta=6.67$ $am=0.007,0.05$ on $20^3\times64$ lattices.}
\label{b676ccorr}
\end{figure}

\begin{figure}[htb]
\rotatebox{270}{\includegraphics[width=15pc]{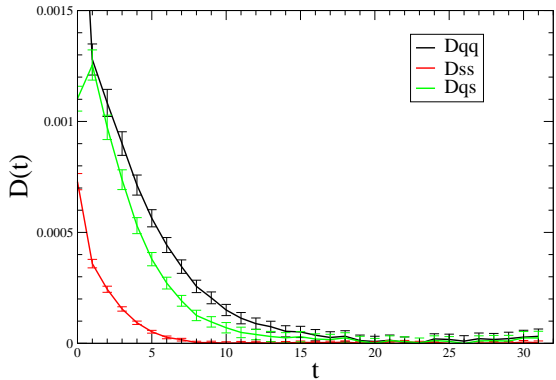}}
\caption{Disconnected correlators of the $\gamma_5\otimes{\bf 1}$ operator
for $N_f=2+1$ $\beta=6.67$ $am=0.007,0.05$ on $20^3\times64$ lattices.}
\label{b676dcorr}
\end{figure}

Our results to date show a clear signal for connected and disconnected
correlators on $N_f=0,2$, and $2+1$ lattices. As an example of connected
correlators, see Figure \ref{b676ccorr}. The corresponding disconnected
correlators are in  Figure \ref{b676dcorr}.

\begin{figure}[htb]
\rotatebox{270}{\includegraphics[width=15pc]{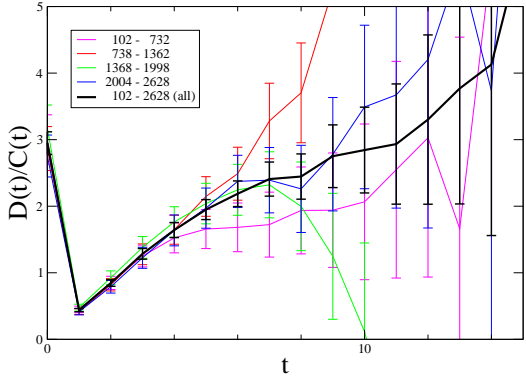}}
\caption{Full $D/C$ ratio for $\gamma_5\otimes{\bf 1}$ operator
for $N_f=2+1$ $\beta=6.67$ $am=0.007,0.05$ on $20^3\times64$ lattices. Bold
curve represents data from the full ensemble, while the finer grey curves 
correspond to bins of one quarter of the ensemble.}
\label{b676rat}
\end{figure}

To compute the $D/C$ ratio for the $N_f=2+1$ case, we must generalize 
the definition, (\ref{full_ratio}), of $D/C$. For non-degenerate flavors
where the $N_f$ total flavors have been separated into $N_q$ light flavors
and $N_s$ strange flavors we must replace:
\begin{eqnarray}
N_f^2D&\longrightarrow& N_q^2D_{qq} + N_s^2D_{ss} + 2N_qN_sD_{qs}\\
N_fC &\longrightarrow&N_qC_{qq} + N_s C_{ss}.
\end{eqnarray}
Here $D_{qq}$ and $D_{ss}$ are light-light and strange-strange disconnected 
correlators respectively. $D_{qs}$ is the mixed correlator, with a light
quark operator correlated with a strange quark operator. Similarly $C_{qq}$
and $C_{ss}$ are light-light and strange-strange connected correlators 
respectively. So the $D/C$ ratio generalized for $2+1$ flavors is
\begin{equation}
R(t)=\frac{4D_{qq} + D_{ss}+ 4D_{qs}}{4C_{qq} + C_{ss}}.
\label{full_ratio21}
\end{equation}
For each ensemble we computed $R(t)$. We show an example of $R(t)$ for 
$N_f=2+1$ flavors in Figure \ref{b676rat}. It is here that the limitations 
of our current statistics begin to show. When we divide the data set into bins 
of approximately 100 configurations each we note significant differences 
between the bins in the behavior of $R(t)$ for $t>6$. We attribute this to 
gauge configuration noise in the disconnected correlator. It suggests
that accurate determination of the dynamical staggered pseudoscalar singlet
propagator will likely come only with data sets with significantly more 
configurations. Additional the behavior of $R(t)$ indicates that there 
may be an inconsistency with the relative normalizations of our connected
and disconnected correlators. 

\subsection{Variance reduction tests}
\label{var_red}
\begin{figure}[htb]
\rotatebox{270}{\includegraphics[width=15pc]{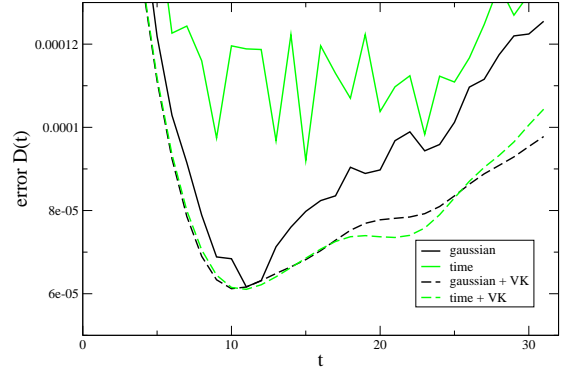}}
\caption{Comparison of disconnected correlator errors using 
stochastic sources with time-slice dilution and the Venkataraman-Kilcup (VK)
trick on 36 $N_f=2+1$ $\beta=6.67$ $am=0.01,0.05$ on $20^3\times64$
configurations.}
\label{b676dcor_err_comp}
\end{figure}

Faced with the the large uncertainties of the disconnected propagator 
measures in the $N_f=2+1$ cases, we tested several several variance 
reduction schemes to try to improve the signal-to-noise ratio of our 
disconnected correlators. We ran our disconnected correlator measurement 
routine on a set of 36 lattices, using a number of different noise vector 
dilution schemes as well as the Venkataraman-Kilcup variance 
reduction (VKVR) trick.
A comparison is in 
Figure \ref{b676dcor_err_comp}. 

There is no doubt that Venkataraman-Kilcup
variance reduction is advantageous. In contrast to reference 
\cite{Foley:2005ac}, we found no apparent advantages to any of the 
dilution schemes alone, although it may be possible that with such a small 
number of lattices, gauge noise overwhelmed the reduction in source noise.
There is some
suggestion of a slight advantage when time-slice dilution was combined 
with VKVR. Figure \ref{b676dcor_err_comp} shows a comparison of disconnected 
correlator error with and without VKVR and time-slice dilution.

It is 
difficult to make definitive statements without careful error analysis of the 
variance.
The volume filling noise sources can be swapped for time-slice-diluted noise 
for no extra computational cost and
disconnected operators can be computed both with and without VKVR at no 
extra cost. So both are implemented in our in-progress $N_f=2$ and $N_F=0$ 
runs, and will be used in future $N_f=2+1$ runs.

\section{CONCLUSIONS \& OUTLOOK}
We present this as a status report of our staggered pseudoscalar singlet
physics project, which is very much a work in progress. We have measured
unambiguous signals for connected and disconnected correlators persisting
through as many as a dozen time-slices. The gauge error inherent in the 
disconnected correlators has limited our ability to precisely determine the 
$D/C$ ratio, and it appears that significantly longer timeseries will be 
necessary.

In part to meet the significant challenges of measuring disconnected 
correlators for singlet physics, UKQCD has begun generating the first of 
several long-timeseries dynamical Asqtad fermion ensembles on the QCDOC
machine. Figure \ref{qcdoc_configs} lists the lattice sizes and timeseries
lengths for the planned ensembles.  With these larger ensembles gauge noise 
should be controlled well enough to make high-precision measurements of the 
disconnected correlators. At that point it should be possible to not only 
scrutinize the behavior of the $D/C$ ratio, but variational fitting of the
full pseudoscalar singlet propagator should allow determination of 
both the $\eta$ and $\eta'$ meson masses for $N_f=2+1$ flavors of dynamical
staggered lattice fermions.

\begin{table}[htb]
\label{qcdoc_configs}
\begin{tabular}{@{}llll}
$a$ (fm) & $m_q/m_s$ & $L^3\times T$ & Trajectories\\
\hline
0.125 & 0.2 & $24^3\times 64$ & 30000\\
0.09 & 0.2 & $32^3\times 96$ & 20000\\
0.06 & 0.2 & $48^3\times 144$ & 2000\\
\end{tabular}\\[2pt]
\caption{Asqtad staggered ensembles planned for generation on the UKQCD's QCDOC.}
\end{table}


\begin{thebibliography}{9}
\bibitem{Witten:1979vv}
  E.~Witten,
  Nucl.\ Phys.\ B  156 (1979) 269.

\bibitem{Veneziano:1979ec}
  G.~Veneziano,
  Nucl.\ Phys.\ B   159 (1979) 213.

\bibitem{Itoh:1987iy}
  S.~Itoh, Y.~Iwasaki and T.~Yoshie,
  Phys.\ Rev.\ D  36 (1987) 527.

\bibitem{McNeile:2000hf}
  C.~McNeile and C.~Michael, 
  Phys.\ Lett.\ B  491 (2000) 123,
  [Erratum-ibid.\ B  551 (2003) 391].

\bibitem{Struckmann:2000bt}
  T.~Struckmann {\it et al.}, 
  Phys.\ Rev.\ D  63 (2001) 074503.

\bibitem{Lesk:2002gd}
  V.~I.~Lesk {\it et al.}, 
  Phys.\ Rev.\ D 67 (2003) 074503.



\bibitem{Schilling:2004kg}
  K.~Schilling, H.~Neff and T.~Lippert,
Lect.\ Notes Phys.\   663 (2005) 147. 

\bibitem{Allton:2004qq}
  C.~R.~Allton {\it et al.}, 
  Phys.\ Rev.\ D  70 (2004) 014501.


\bibitem{DeGrand:2002gm}
  T.~DeGrand and U.~M.~Heller, 
 Phys.\ Rev.\ D  65 (2002) 114501.


\bibitem{Venkataraman:1997xi}
  L.~Venkataraman and G.~Kilcup,
  [ arXiv:hep-lat/9711006].


\bibitem{Kogut:1998rh}
  J.~B.~Kogut, J.~F.~Lagae and D.~K.~Sinclair,
  Phys.\ Rev.\ D  58 (1998)  054504.

\bibitem{Bernard:1992mk}
  C.~W.~Bernard and M.~F.~L.~Golterman,
  Phys.\ Rev.\ D  46 (1992) 853.

\bibitem{Bernard:1993sv}
  C.~W.~Bernard and M.~F.~L.~Golterman,
  Phys.\ Rev.\ D  49 (1994) 486.

\bibitem{Bernard:2001av}
  C.~W.~Bernard {\it et al.},
Phys.\ Rev.\ D  64 (2001) 054506.

\bibitem{Aubin:2004wf}
  C.~Aubin {\it et al.},
  Phys.\ Rev.\ D  70  (2004) 094505.

\bibitem{Orginos:1998ue}
  K.~Orginos and D.~Toussaint, 
  Phys.\ Rev.\ D  59 (1999) 014501.


\bibitem{Orginos:1999cr}
  K.~Orginos, D.~Toussaint and R.~L.~Sugar, 
  Phys.\ Rev.\ D  60 (1999)  054503.

\bibitem{Lepage:1998vj}
  G.~P.~Lepage,
  Phys.\ Rev.\ D  59 (1999) 074502.

\bibitem{Edwards:2004sx}
  R.~G.~Edwards and B.~Joo, 
  Nucl.\ Phys.\ Proc.\ Suppl.\  140 (2005) 832.

\bibitem{Bitar:1988bb}
  K.~Bitar, {\it et al.},
  Nucl.\ Phys.\ B 313 (1989) 348. 


\bibitem{Fiebig:1990uh}
  H.~R.~Fiebig and R.~M.~Woloshyn,
  Phys.\ Rev.\ D  42 (1990) 3520.


\bibitem{Dong:1993pk}
  S.~J.~Dong and K.~F.~Liu,
  Phys.\ Lett.\ B 328 (1994) 130.

\bibitem{Kuramashi:1994aj}
  Y.~Kuramashi, {\it et al.},
  Phys.\ Rev.\ Lett.\  72 (1994) 3448.

\bibitem{Eicker:1996gk}
  N.~Eicker, {\it et al.}, 
  Phys.\ Lett.\ B 389 (1996) 720.


\bibitem{Foley:2005ac}
  J.~Foley, {\it et al.},
  arXiv:hep-lat/0505023 (2005).


\end{thebibliography}
\end{document}